# Complex magnetic order in nickelate slabs

M. Hepting,[1,*] R. J. Green,[2,3] Z. Zhong,[1,†] M. Bluschke,[4,5] Y. E. Suyolcu,[1] S. Macke,[1] A. Frano,[6,‡] S. Catalano,[7] M. Gibert,[7,§] R. Sutarto,[8] F. He,[8] G. Cristiani,[1] G. Logvenov,[1] Y. Wang,[1] P. A. van Aken,[1] P. Hansmann,[1,9] M. Le Tacon,[1,¶] J.-M. Triscone,[7] G. A. Sawatzky,[2] B. Keimer,[1] and E. Benckiser[1,**]

[1]*Max Planck Institute for Solid State Research, Heisenbergstr. 1, 70569 Stuttgart, Germany*
[2]*Quantum Matter Institute, Department of Physics and Astronomy, University of British Columbia, Vancouver, British Columbia V6T 1Z1, Canada*
[3]*Department of Physics & Engineering Physics, University of Saskatchewan, Saskatoon, Saskatchewan, Canada S7N 5E2*
[4]*Max Planck Institute for Solid State Research, Heisenbergstr. 1, 70569 Stuttgart*
[5]*Helmholtz-Zentrum Berlin für Materialien und Energie, Wilhelm-Conrad-Röntgen-Campus BESSY II, Albert-Einstein-Str. 15, D-12489 Berlin, Germany*
[6]*Department of Physics, University of California, Berkeley, California 94720, USA*
[7]*Department of Quantum Matter Physics, University of Geneva, CH-1211 Geneva, Switzerland*
[8]*Canadian Light Source, Saskatoon, Saskatchewan S7N 2V3, Canada*
[9]*Institut für Theoretische Physik, Eberhard Karls Universität Tübingen, Auf der Morgenstelle 14, 72076 Tübingen, Germany*

**Magnetic ordering phenomena have a profound influence on the macroscopic properties of correlated-electron materials, but their realistic prediction remains a formidable challenge. An archetypical example is the ternary nickel oxide system $R$NiO$_3$ ($R$ = rare earth), where the period-four magnetic order with proposals of collinear and non-collinear structures and the amplitude of magnetic moments on different Ni sublattices have been subjects of debate for decades.[1–6] Here we introduce an elementary model system – NdNiO$_3$ slabs embedded in a non-magnetic NdGaO$_3$ matrix – and use polarized resonant x-ray scattering (RXS) to show that both collinear and non-collinear magnetic structures can be realized, depending on the slab thickness. The crossover between both spin structures is correctly predicted by density functional theory and can be qualitatively understood in a low-energy spin model. We further demonstrate that the amplitude ratio of magnetic moments in neighboring NiO$_6$ octahedra can be accurately determined by RXS in combination with a correlated double cluster model. Targeted synthesis of model systems with controlled thickness and synergistic application of polarized RXS and *ab-initio* theory thus provide new perspectives for research on complex magnetism, in analogy to two-dimensional materials created by exfoliation.[7]**

Recent progress in the synthesis of epitaxial heterostructures and thin films has opened new opportunities for designing and manipulating correlated-electron systems. Rare-earth nickel oxides $R$NiO$_3$ with pseudocubic perovskite structure are a prominent example. In bulk form, $R$NiO$_3$ with $R \neq$ La show temperature driven metal-insulator ($T_{MI}$) and Néel ($T_N$) transitions with $T_{MI} \geq T_N$ (ref. 8). While the metal-insulator transition is understood in terms of charge/bond disproportionation,[9–13] the nature of the antiferromagnetic ground state remains unresolved. Its unusually large periodicity is very robust and characterized by the wave vector $\mathbf{q}_0 = (1/4, 1/4, 1/4)$ in pseudocubic notation with proposals of collinear (↑↑↓↓) and non-collinear (↑→↓←) spin arrangements.[1,2,4,5,10,14,15] Owing to the difficulty of synthesizing large single crystals, most recent insights were gained from work on epitaxial thin films or multilayers. It has been shown that $\mathbf{q}_0$ magnetic order can be induced by confinement of the electron system[6,16] and can occur with or without concomitant bond disproportionation.[17,18] Furthermore the orientation of magnetic moments in the period-four order relative to the crystallographic axes can be systematically influenced in heterostructures, e.g. by the epitaxial relation to the underlying substrate[6] or interfacial interactions.[19,20] Theoretical descriptions of the unusual magnetic order vary from localized spin models with antisymmetric exchange interactions to itinerant models[10] where the spin order arises from a nesting instability of the Fermi surface. Diverse collinear and non-collinear ground states have been discussed.[15] It is thus of great importance to devise new experimental tests of theoretical concepts describing the unusual magnetism of the nickelates and its interplay with the bond-order instability.[21,22]

In the present work we introduce NdNiO$_3$ slabs oriented along the magnetic propagation vector $\mathbf{q}_0$ as a specifically designed magnetic model system where the spin configuration and the magnetic moment amplitudes can be manipulated and tested against theory. The slabs were realized in thin film structures with exceptionally high quality, exhibiting virtually no defects over lateral length scales of ∼ 100 nm

*Present address: Stanford Institute for Materials and Energy Sciences, SLAC National Accelerator Laboratory and Stanford University, 2575 Sand Hill Road, Menlo Park, California 94025, USA

†Present address: Key Laboratory of Magnetic Materials and Devices & Zhejiang Province Key Laboratory of Magnetic Materials and Application Technology, Ningbo Institute of Materials Technology and Engineering, Chinese Academy of Sciences, Ningbo 315201, China

‡Present address: Department of Physics, University of California San Diego, La Jolla, California 92093, USA

§Present address: Physik-Institut, Univeristy of Zurich, Winterthurerstrasse 190, 8057 Zürich, Switzerland

¶Present address: Karlsruher Institut für Technologie, Institut für Festkörperphysik, Hermann-v.-Helmholtz-Platz 1, D-76344 Eggenstein-Leopoldshafen

**E.Benckiser@fkf.mpg.de

(see the Methods section). Polarized RXS is used in conjunction with recent model calculations[23] for an accurate determination of the magnetic ground state. With decreasing layer thickness we track an unexpected crossover from non-collinear (↑→↓←) to collinear (↑↑↓↓) antiferromagnetic order as the nickelate structure is truncated to two magnetic periods or less. Furthermore, we extract the magnitudes of the magnetic moments in the Ni sublattices. For thicker bulk-like slabs we find unequal moments in agreement with RXS experiments on $NdNiO_3$ films[18,24] but contrary to neutron powder diffraction studies which found nearly equal moment sizes in $NdNiO_3$.[2] In thinner embedded slabs we find the disparity in the moment amplitudes to be reduced due to the higher crystal symmetry of the embedding material. The approach of cropping magnetic exchange paths along specifically chosen crystallographic directions and the concomitant RXS methodology we introduce are widely applicable to research on complex magnetism.

We have synthesized $NdNiO_3$ in [111] oriented slabs comprising $N = 55$, 45, 14, and 8 consecutive Ni layers sandwiched between layers of $NdGaO_3$ [Fig. 1(a),(b)]. In bulk $NdNiO_3$, $T_{MI}$ is accompanied by a structural phase transition from orthorhombic to monoclinic symmetry ($Pbnm \rightarrow P2_1/n$).[25] In the monoclinic phase the $NiO_6$ octahedra split into sets of expanded and contracted octahedra with long Ni-O bonds (LB) and short bonds (SB), arranged in the typical breathing mode pattern and yielding alternating LB and SB planes along the pseudocubic [111] direction [Fig. 1(b)]. Previously, nickelate heterostructures were investigated with the intention to manipulate the $R$NiO$_3$ magnetism extrinsically via the interplay with a second magnetic material,[19] via charge transfer to the $Ni^{3+}$ ions,[26] or via an altered orbital occupation.[6] In contrast, here we aim to explore the pure effect of truncation of magnetic exchange bonds along the [111] direction and its effect on the period-four magnetic order to gain information on the relevant magnetic exchange interactions in this material. Consequently we have chosen to stabilize the thin $NdNiO_3$ slabs through $NdGaO_3$, a well characterized paramagnetic band insulator with a stable valence $Ga^{3+}$ ion, preventing a polar discontinuity at the interface and charge transfer to the $NdNiO_3$. Furthermore, the $R$-cation sublattice remains unchanged across the $NdGaO_3$-$NdNiO_3$ interface precluding additional symmetry breaking and intermixing. Moreover, both materials show excellent epitaxial matching conditions along the [111] direction with closely similar oxygen octahedra tilt angles.[27] This is of particular importance since distinct changes in the tilt angles can facilitate a realignment of the easy axis in magnetic systems.[28] The high crystalline quality of our samples was confirmed by scanning transmission electron microscopy (STEM) high angle annular dark field (HAADF) [Fig. 1(c)] and annular bright field (ABF) imaging, together with x-ray reflectivity characterization (see Methods). The chemical sharpness of the heterointerfaces is verified by electron energy loss spectroscopy (EELS) [Fig. 1(c)]. Finally, the [111] orientation is expected to prevent strain induced orbital polarization of the Ni $e_g$ orbitals [Fig. 2(a)], which was identified as the origin of reorientation of the spin spiral polarization plane in strained [001] heterostructures via the spin orbit coupling[6] and would obscure the observation of the elementary effect of [111] directional truncation.

The resonant x-ray scattering (RXS) experiments were performed with $\pi$ and $\sigma$ polarized photons and energies tuned to the Ni $L_3$ edge (see Methods). We observed the antiferromagnetic (1/4, 1/4, 1/4) Bragg reflection, with decreasing $T_{\mathbf{N}}$ for decreasing $NdNiO_3$ layer thickness $N$, presumably as a consequence of the reduced dimensionality of the spin system [Fig. 2(b),(c) and Methods]. The intensity of magnetic scattering depends on the relative orientation between the magnetic moments and the polarization of the incoming light (see Methods). A rotation around the azimuthal angle $\psi$ systematically varies this relative orientation, and the $\psi$-dependence of the scattering for the $\boldsymbol{\pi} \equiv I_{\pi\pi} + I_{\pi\sigma}$ and $\boldsymbol{\sigma} \equiv I_{\sigma\pi}$ channels can be modeled following the formalisms described in refs. 6. Our results together with the model fits are summarized in Fig. 3. Panel 3(a) shows as reference the simulated $\boldsymbol{\pi}$ and $\boldsymbol{\sigma}$ intensities for the fully orthogonal spiral [Fig. 3(g)]. The shaded area corresponds to a breathing distortion with bond-length difference $\delta d$ ranging from 0 to 0.037 Å.[25] Its relevance will be discussed below. Two characteristic features of the azimuthal dependence are the nodes at $\psi = 90°$ and 270° and the maxima of $\boldsymbol{\pi}$ and minima of $\boldsymbol{\sigma}$ at $\psi = 0°$ and 180°. The simulated curves for collinear spin order along the [111] direction (dashed lines) predict a constant scattering signal as a function of $\psi$ for our measurement geometry. The experimental data of the $N = 55$ sample in Fig. 3(b) show a shift of the first node to $\psi = 60°$ with different $\boldsymbol{\pi}$ and $\boldsymbol{\sigma}$ intensities, while a crossing of $\boldsymbol{\pi}$ and $\boldsymbol{\sigma}$ occurs at the second node (240°). From our model fits we find that (i) a shift of the whole pattern in $\psi$ is initiated by a spiraling of a spin sublattice within the (111) plane and (ii) the crossing and separation of the nodes comes from spin reorientation out of the (111) plane [Fig. 3(h), orange spins]. The full set of numerical values for the obtained spin angles can be found in the Supplementary Information. For the sample with $N = 45$ [Fig. 3(c),(i)] the entire azimuthal pattern is shifted to slightly lower $\psi$ angles as the second spin sublattice spirals further within the (111) plane. Notable changes occur for the $N = 14$ sample, with a strongly reduced overall modulation amplitude of $\boldsymbol{\pi}$ and $\boldsymbol{\sigma}$ and the second spin sublattice increasingly misaligned to the (111) planes. For the thinnest sample with $N = 8$ the azimuthal dependence has drastically changed such that the amplitude of the modulation with $\psi$ is almost zero. [Fig. 3(e)]. The symmetry of the modulation is a signature of collinear spin arrangement and the small modulation amplitude is indicative of an alignment close to the [111] direction [Fig. 3(k)]. Alternative model fits are discussed in the Supplementary Information, however, we emphasize that the central statement of a change of the spin configuration from non-collinear to collinear is reproduced in all models.

Slabs with variable thickness and truncation along specific crystallographic directions represent an interesting new testing ground for *ab-initio* theories. Motivated by the experimentally observed crossover between collinear and non-collinear magnetism, we have performed DFT calculations with an on-site Hubbard $U$ for fully relaxed cubic and constrained monoclinic structures in bulk and finite-thickness slab geometries [Fig. 4(a)-(e) and Methods]. When comparing the ground state energies of different magnetic configurations in cubic symmetry, the non-collinear period-four order is found to be the lowest-energy state in bulk $NdNiO_3$ [Fig. 4(a)], in agreement with experiment (ref. 4 and Supplementary Information). Remarkably, the non-collinear orthogonal state is disfavored for [111] directional truncation of $NdNiO_3$ with $N = 4$ Ni layers and a collinear antiferromagnetic configuration is found to be the ground state [Fig. 4(b)]. When increasing the number of layers to $N = 8$, predominantly collinear order is still observed, albeit with a trend of spin canting in the slab center. This trend increases with slab thickness, and for $N = 16$, the slab center already exhibits the orthogonal non-collinear configuration characteristic of bulk $NdNiO_3$ [Fig. 4(d)]. When calculating the mean angle between neighboring spin axes we obtain good agreement with the experimentally determined angles between the two sublattices [Fig. 4(e), left].

The results of our first-principles calculations thus reproduce the slab-thickness-induced crossover from collinear to spiral magnetism, in excellent agreement with our experimental observations (Fig. 3). In order to gain insight into the driving mechanism, we consider the effect of truncation of exchange paths in a low-energy spin model. Recent experimental and theoretical work on transition metal oxides close to metal-insulator transitions suggested that the spin ordering pattern follows from competing nearest ($J_1$) and next-nearest ($J_2$) exchange interactions, and that the mechanism that stabilizes the 90° mutual spin orientation originates from frustration due to a double exchange (DE) term in $J_1$ arising from low-energy charge fluctuations.[29] The DE process depends on the angle $\alpha$ between spins[30] at the LB and SB sites and the energy gain for each bond is $-\cos\frac{\alpha}{2}$, i.e. proportional to $-(N-1)\cos\frac{\alpha}{2}$ for a slab of $N$ magnetic layers. Accordingly the energy gain for mutual 90° alignment (↑→↓←) is $-(N-1)/\sqrt{2}$ while the energy gain from alternatingly 0° and 180° aligned spins along [111] (↑↑↓↓) is $-N/2$. The ratio of these energy gains is plotted in Fig. 4(e) as a function of the number of magnetic layers $N$ and compared to the angles between SB and LB sublattice moments obtained from experiment and the DFT calculations. While for large $N$ the spiral order is clearly favored, the energy gain relative to the collinear order significantly decreases at small $N$ and the collinear order is further enhanced through additional effects such as the suppression of bond order (and therewith the DE term), as observed in experiment and discussed in the following.

In addition to the changes in the magnetic moment orientation we extract the Ni magnetic moment amplitudes from the azimuthal dependences of Fig. 3. Contrary to previous RXS experiments,[4–6] a systematic anomaly in the energy dependence of the (1/4, 1/4, 1/4) peak for particular $\psi$ angles is observed, reflected in a non-constant $\pi/\sigma$ ratio across the $L_3$ energy range [Fig. 5(a),(b) and insets]. We interpret the anomaly as the unique signature of scattering from two inequivalent Ni sites[18,24] with distinct magnetic scattering tensors, which requires an advanced modeling of the magnetic form factors $f_{\mathrm{LB,\,SB}}$ of the LB and SB sites. To this end, we use a recently introduced double cluster model,[23] which is based on the negative charge transfer concept that starts from a Ni $3d^8$ electron configuration with a ligand hole at the oxygen, as introduced in ref. 31 and implemented in *ab-initio* calculations in refs. 9,11. The double cluster explicitly incorporates the $NiO_6$ breathing distortion whose magnitude $\delta d$ corresponds to the deviation of the long and short Ni-O bond lengths from the mean value. The fits in Fig. 3(b),(c) yield $\delta d = 0.041$ Å for the $N = 55$ and 45 sample which corresponds to magnetic moments of $m_1 = 0.80\ \mu_B$ and $m_2 = 1.44\ \mu_B$ for the SB and LB sites, respectively.[23] The fits for the thinner $NdNiO_3$ slabs [Fig. 3(d),(e)] yield a reduction of the breathing distortion with decreasing $N$, that is $\delta d = 0.025$ Å ($m_1 = 0.96\ \mu_B$, $m_2 = 1.28\ \mu_B$), and $\delta d = 0.002$ Å ($m_1 = 1.10\ \mu_B$, $m_2 = 1.14\ \mu_B$) for $N = 14$ and 8, respectively. The equalization of the moment magnitudes for $N = 8$ is in accordance with the absence of the energy scan anomaly in Fig. 5(c) and indicates the presence of only one crystallographic Ni site.

This is further supported by STEM ABF oxygen mapping which reveals a full structural pinning of the $NiO_6$ octahedral network to the confining $NdGaO_3$ layers for $N = 8$, whereas subtle relaxations in the Ni-O-Ni bond angles are observed for the thick $N = 55$ film (see Methods and Supplementary Information). Hence the orthorhombic $NdGaO_3$ symmetry can be imposed on thin $NdNiO_3$ slabs and hamper the emergence of the monoclinic breathing distortion at low temperatures, resulting in antiferromagnetic order with nearly equivalent Ni moment sizes.

To explore the relevance of the lower crystal symmetry in the magnetically ordered bulk phase of NNO (monoclinic $P2_1/n$), we repeated the DFT+$U$ calculation for supercells allowing for such lower-symmetric structures (see Methods). Independent of crystal symmetry, our calculations reproduce the crossover to the collinear configuration for $N \leq 8$ Ni layers. We thus conclude that the structural distortions (octahedral tilts and rotations) and the presence of two different Ni sites are no precondition for the crossover to occur. Furthermore, we found no relevant influence of spin-orbit coupling on the reported results and therefore identify the truncation of magnetic bonds at the interfaces as the leading effect.

To conclude, we have introduced atomically thin slabs embedded into a non-magnetic matrix as a novel platform for the investigation of complex magnetic ordering phenomena. The period-four antiferromagnetic order in $NdNiO_3$ is found to be very robust and even present in slabs which are only two magnetic periods thick. Furthermore its appearance is revealed to be independent of concomitant static bond disproportionation. The observed manifestation of two types of spin order, non-collinear and collinear, within the same material system yields deep insights into the underlying magnetic exchange interactions prevalent in the nickelates, and is reproduced by *ab-initio* theory and qualitatively explained through the combined effects of truncation and suppressed bond order in a low-energy spin model.

Looking beyond the nickelates, the targeted modification of magnetic exchange bonds along a particular crystallographic direction in a well-defined, highly-ordered heterostructure is a powerful concept to gain deep insights into the mechanisms underlying complex magnetic order. In particular, in materials with different and possibly competing magnetic interactions, the approach we have presented opens new perspectives for research on critical behavior at boundaries and in low dimensions, and for the systematic exploration and manipulation of complex spin textures such as magnetic vortices and skyrmions.

## Acknowledgements

We thank G. Khaliullin, I. Elfimov, Y. Lu, C. Dietl, F. Wrobel, H.-U. Habermeier, and P. Wochner for the fruitful discussions. Financial support from the DFG under Grant No. SFB/TRR80 G1 and from the European Union Seventh Framework Program [FP/2007-2013] under grant agreement No. 312483 (ESTEEM2) is acknowledged. Part of this work has been funded by the Max Planck-UBC Centre for Quantum Materials. Further, this work was supported by the Swiss National Science Foundation through Division II. The research leading to these results has received funding from the European Research Council under the European Union's Seventh Framework Program (FP7/2007-2013)/ERC Grant Agreement no. 319286 (Q-MAC). The Canadian Light Source (CLS) is funded by the Canada Foundation for Innovation, NSERC, the National Research Council of Canada, the Canadian Institutes of Health Research, the Government of Saskatchewan, Western Economic Diversification Canada, and the University of Saskatchewan.

## Author contributions

M.H., R.J.G., and E.B. conceived the project, performed the experiments, and analyzed the data together with M.L.T., G.A.S. and B.K. Assistance in the experiments and contributions to the data analysis were made by M.B., S.M. and A.F. STEM investigations were performed by Y.E.S. under the supervision of Y.W. and P.A.v.A. The DFT+$U$ calculations were carried out by Z.Z. and P.H. The RXS experiments were supported by R.S. and F.H. The PLD samples were grown by G.C. and G.L. The sputtered sample was grown and characterized by S.C. under the supervision of M.G. and J-M.T. M.H., B.K., and E.B. wrote the manuscript with contributions from all authors.

## Additional information

The authors declare no competing financial interests. Supplementary information accompanies this paper on www.nature.com. Correspondence and requests for materials should be addressed to E.B.

---

(2000).

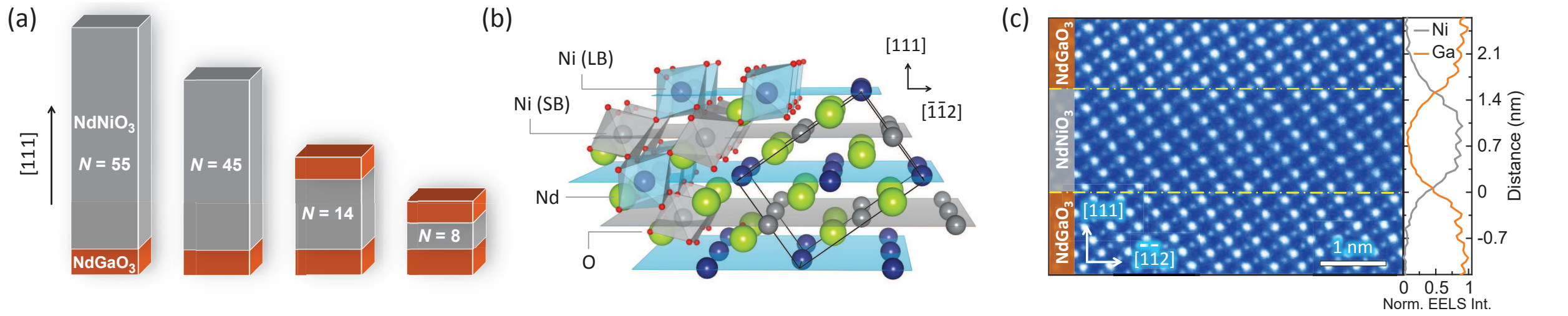

Fig. 1: **$NdNiO_3$ slabs truncated along the [111] crystallographic direction.** (a) Schematic of the $NdNiO_3$ slabs grown on $NdGaO_3$ substrate with $N$ indicating the number of Ni planes along the pseudocubic [111] direction. The thin $N = 14$ and 8 samples are additionally capped with $NdGaO_3$. The [111] direction coincides with the $\mathbf{q}_0 = (1/4, 1/4, 1/4)$ magnetic propagation vector. (b) Bulk $NdNiO_3$ in [111] orientation. Planes of long-bond (LB, blue) and short-bond (SB, gray) $NiO_6$ octahedra are stacked alternately. The monoclinic $P2_1/n$ unit cell is indicated by black lines. (c) STEM-HAADF image of the $N = 8$ sample (left). Ni and Ga $L_{3,2}$ EELS line scan profiles (right) show the chemical sharpness of the $NdNiO_3$-$NdGaO_3$ interfaces with Ni and Ga intermixing limited to $\sim 2$ atomic layers.

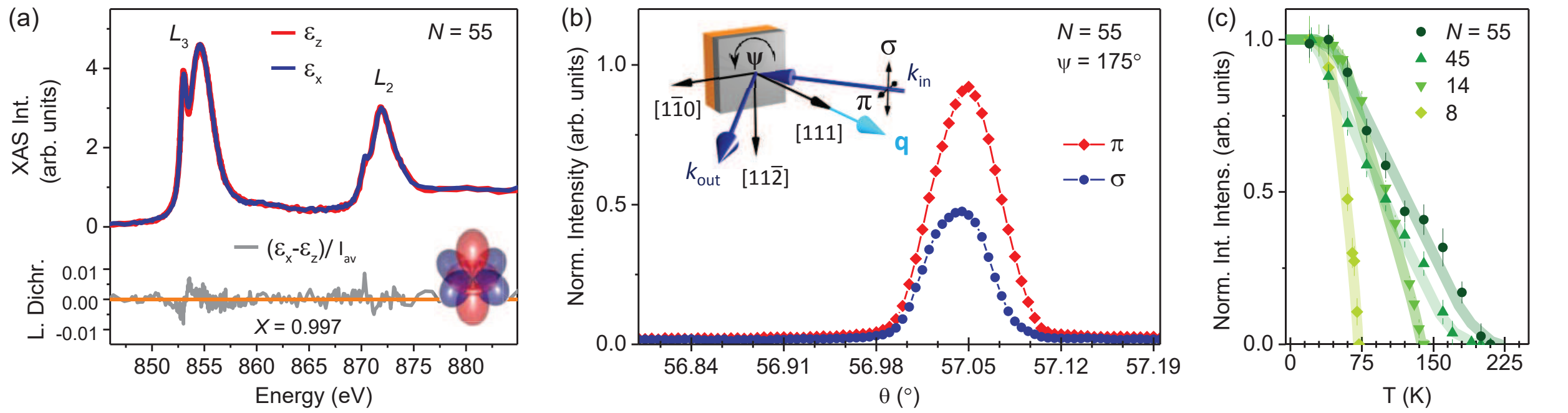

Fig. 2: **X-ray absorption and resonant magnetic x-ray scattering.** (a) X-ray absorption spectra across the Ni $L_{3,2}$ edges at $T = 20$ K measured in tilted wedge geometry with the incoming photon polarizations $\varepsilon_x$ and $\varepsilon_z$ probing the holes in the $e_g$ orbitals of $x^2$-$y^2$ and $3z^2$-$r^2$ symmetry (see inset), respectively. The linear dichroism is defined as $\varepsilon_x$-$\varepsilon_z$ and normalized by the averaged energy integral from 850 to 880 eV. The calculated ratio of holes $X$ in the $e_g$ orbitals is close to unity, indicating the absence of orbital polarization (see the Supplementary Information for details). (b) Rocking curve of the antiferromagnetic (1/4, 1/4, 1/4) Bragg peak measured with incoming photon energy in resonance to the Ni $L_3$ edge ($E = 853$ eV) in $\pi$ and $\sigma$ polarization, respectively. The inset shows the horizontal diffraction geometry used for the magnetic x-ray scattering. For the azimuthal angle dependence the sample is rotated around the angle $\psi$, with the crystallographic directions [111] and [$1\bar{1}0$] spanning the scattering plane for $\psi = 0°$. Positive rotation around the scattering vector $\mathbf{q}$ is right handed. (c) Temperature dependence of the integrated intensity of the antiferromagnetic Bragg peak. The shaded solid lines are guides to the eye. The error bars are two times the standard error of the least-square fits of the Bragg peaks with Lorentzian profiles.

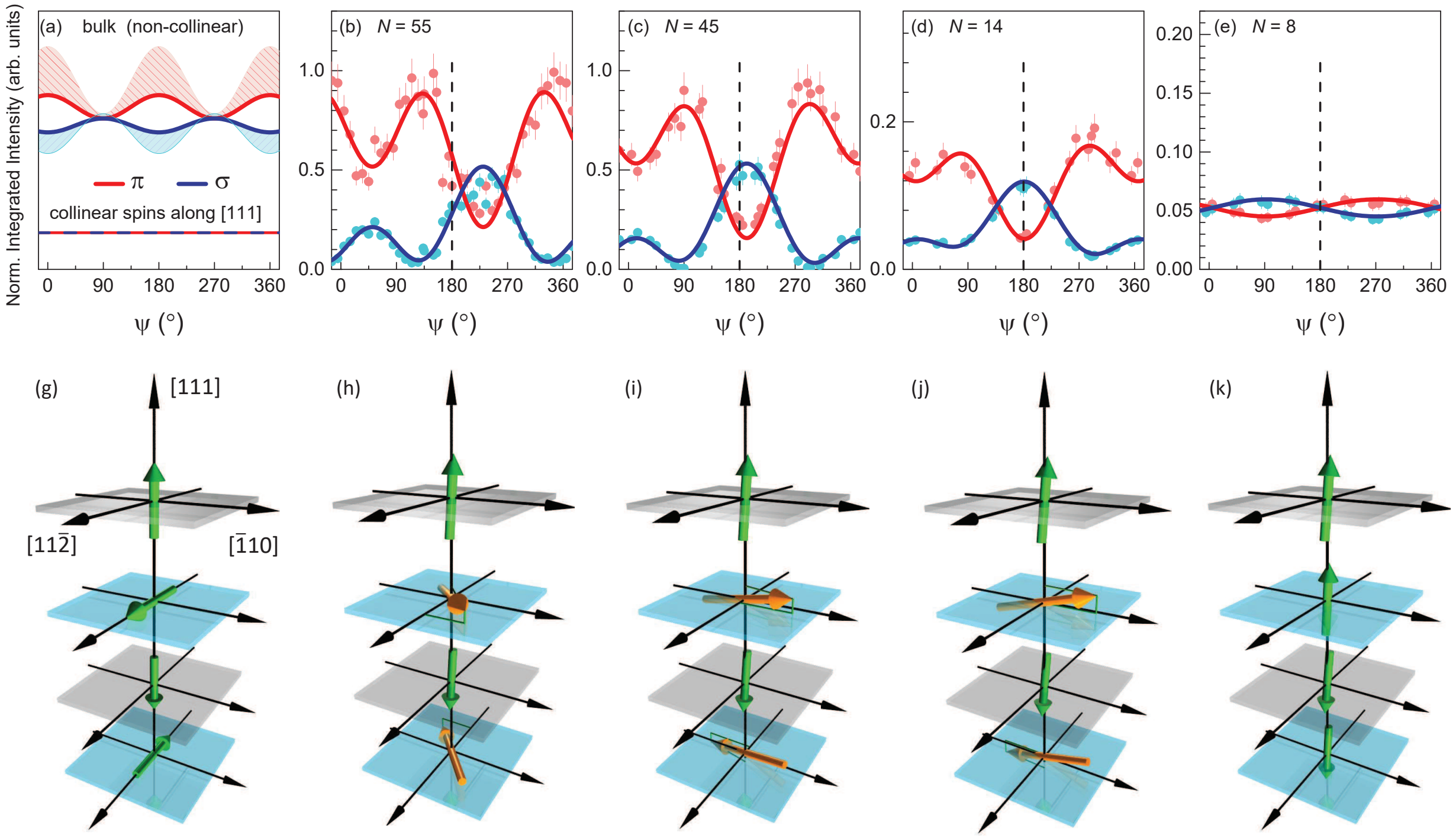

Fig. 3: **Crossover from non-collinear (↑→↓←) to collinear (↑↑↓↓) spin structures by truncation along the [111] direction.** (a) Simulated dependence of the magnetic scattering on the azimuthal angle $\psi$ for non-collinear spins tilted by 90° from one to the other and with breathing distortion $\delta d$ ranging from 0 (solid lines) to 0.037 Å (shaded area). The dashed lines (bottom) simulate spins entirely collinear along the [111] direction yielding no modulation with $\psi$ in the given scattering geometry. The simulated curves are shown on an arbitrary intensity scale. (b)-(e) Normalized experimental data at $T = 20$ K and $E = 853$ eV of the $N = 55$, 45, 14, and 8 Ni samples. The symbols are the integrated intensities of the antiferromagnetic (1/4, 1/4, 1/4) Bragg peaks and the solid lines are the result of fits from the model calculations. Error bars on the symbols correspond to two times the standard error of least-square fits with Lorentzian profiles. (g)-(k) Cartoons of the spin orientation in adjacent Ni planes. (g) is the non-collinear 90° order and (h)-(k) show the evolution from non-collinear to collinear as obtained from the model fits in (b)-(e). For decreasing $NdNiO_3$ thickness the spins (orange) in the second and fourth planes spiral around the [111] axis and exhibit a small out-of plane component. (k) Spins in slabs truncated to 8 Ni layers are collinear (↑↑↓↓) but with a slight misalignment to the [111] axis.

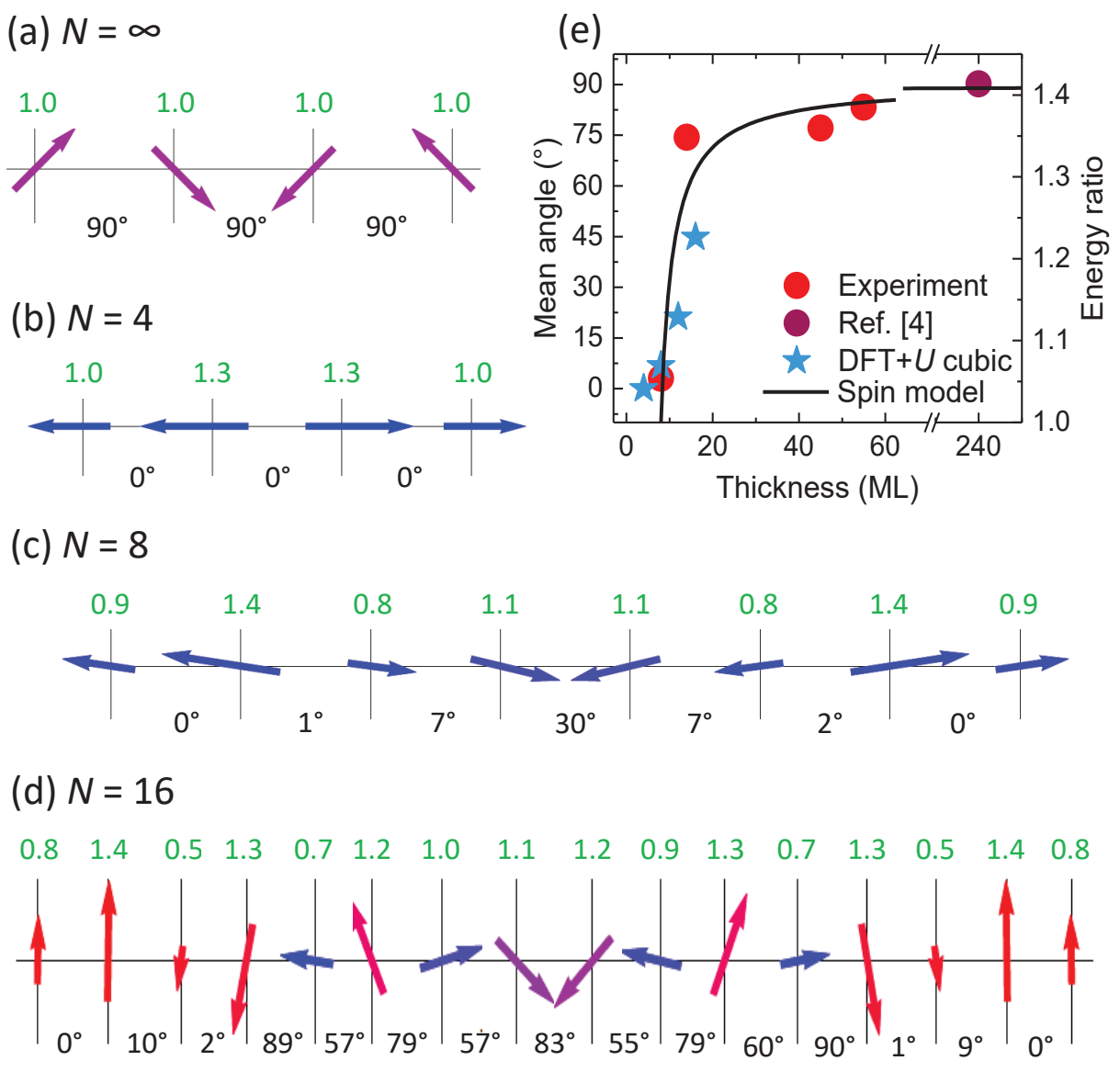

Fig. 4: **DFT+$U$ and low-energy spin model magnetic ground state of $NdNiO_3$-$NdGaO_3$ heterostructures as function of truncation along the [111] direction.** (a)-(d) Representation of the relative orientation and amplitude of the ordered magnetic moments obtained from integration of the site- and spin-resolved density of states for $U$ = 2 eV and cubic crystal symmetry. (a) The unrestricted $N = \infty$ supercell corresponding to bulk $NdNiO_3$ converges to an orthogonal spiral state. (b)-(d) In case of $NdNiO_3$-$NdGaO_3$ slabs truncated to two magnetic periods or less ($N = 8, 4$), the spin configuration converges to the collinear period-four order, while for $N = 16$ the slab center already shows significant deviation from the collinear spin alignment. Experimentally the transition is observed between $N = 14$ and $N = 8$. The green number above each arrow corresponds to its ordered moment amplitude relative to the $N = \infty$ solution and the black number given below describes the angle between the spin axes of neighboring moments. (e) Left: mean angle determined from modeling the azimuthal dependence (Fig. 3 and ref. 4) and by averaging the angles obtained from DFT. Right: ratio of energy gain of the orthogonal spiral ($E_{\uparrow\rightarrow\downarrow\leftarrow}$) compared to the collinear ($E_{\uparrow\uparrow\downarrow\downarrow}$) order obtained from the low-energy spin model (see text) as a function of the number of magnetic layers ($N$). The functional dependence $E_{\uparrow\rightarrow\downarrow\leftarrow}/E_{\uparrow\uparrow\downarrow\downarrow} = \sqrt{2}(1-N^{-1})$ converges to a constant value for $N \rightarrow \infty$.

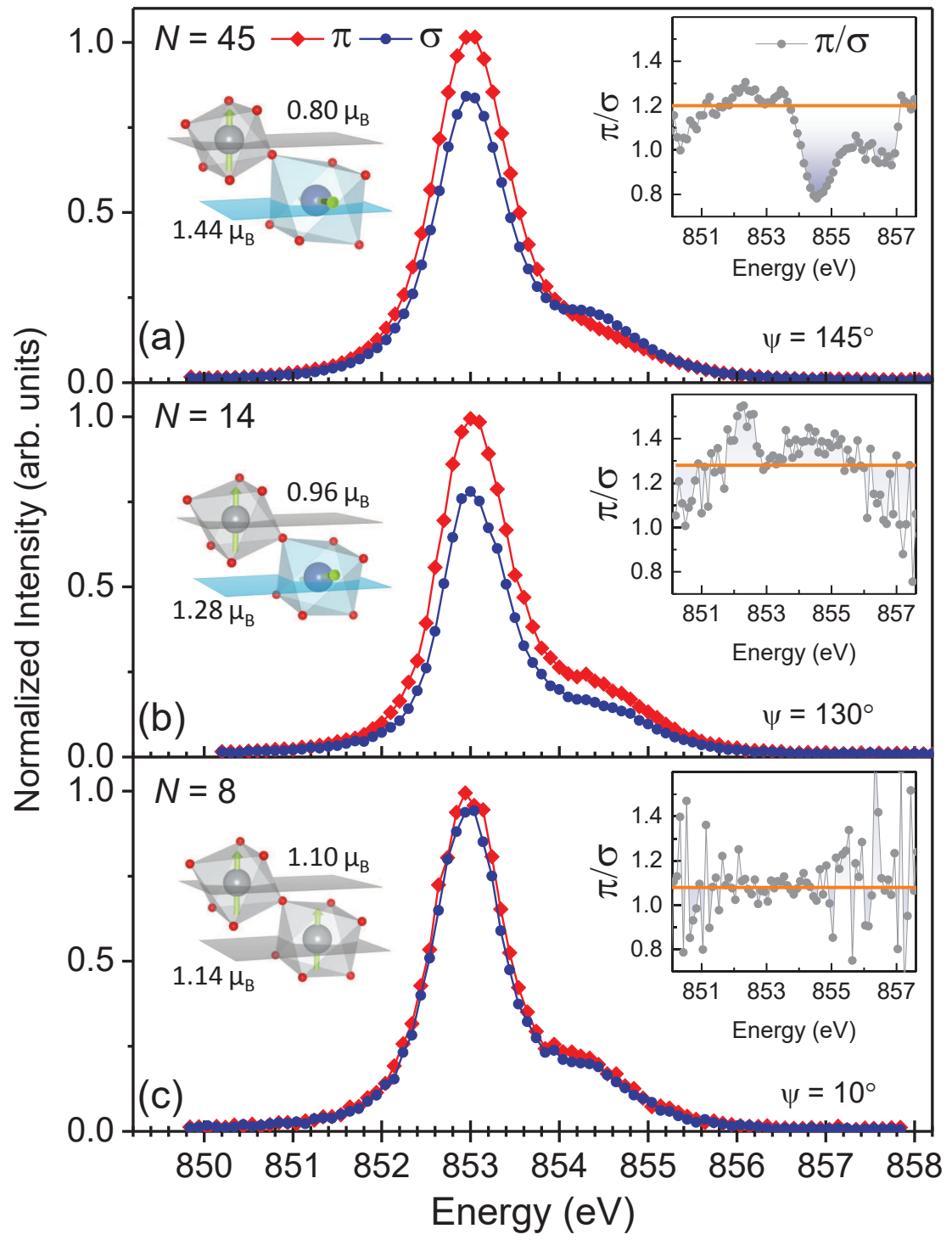

Fig. 5: **Energy dependence of magnetic scattering and tuning of magnetic moments by structural pinning.** (a)-(c) Constant-**q** energy scans of the antiferromagnetic (1/4, 1/4, 1/4) peak with $\pi$ or $\sigma$ polarized x-rays at $T$ = 20 K. (a),(b) The $N$ = 45 and 14 samples show a variation of the $\pi/\sigma$ ratio as a function of the energy (see right insets), indicative of the monoclinic $NiO_6$ breathing distortion and two inequivalent magnetic form factors at the SB and LB Ni sites. The absolute values of the magnetic moments (left insets) can be extracted from a form factor corrected fit to the azimuthal dependence in combination with a double cluster calculation[23] (the error bar is ± 0.08 $\mu_B$). (c) The ultra-thin $N$ = 8 sample does not exhibit the anomaly, suggesting a single set of magnetic form factors and a $NdNiO_3$ crystal structure strongly pinned to the confining orthorhombic $NdGaO_3$, suppressing the $NiO_6$ ground state breathing distortion.

## Methods

**Sample growth:** Samples were grown by pulsed laser deposition[32] ($N = 55$, 14, and 8) and off-axis radiofrequency magnetron sputtering ($N = 45$) on orthorhombic $NdGaO_3$ substrates following the procedures described in refs. 27,32. $NdNiO_3$ layer thicknesses, interface roughnesses, and crystalline quality were determined by on- and off-resonant x-ray diffraction and electron microscopy analysis (see the following paragraphs).

**Resonant x-ray scattering:** RXS experiments were performed at the REIXS beam line of the Canadian Light Source (CLS) with energies tuned to the Ni $L_3$ edge ($2p \rightarrow 3d$). The intensity of magnetic scattering depends on the relative orientation between the magnetic moments and the polarization of the incoming light.[33] Following the formalism in ref. 34 the scattered intensity is

$$I = |\sum_i e^{\imath(\mathbf{k}_{in}-\mathbf{k}_{out})\cdot\mathbf{r}_i}\boldsymbol{\varepsilon}^*_{out}\cdot F_i\cdot\boldsymbol{\varepsilon}_{in}|^2,$$

with $\boldsymbol{\varepsilon}_{in(out)}$ representing the polarization vectors of the electric field of the incoming (outgoing) light, $\mathbf{k}_{in(out)}$ the wave vector of the incoming (outgoing) light, $\mathbf{r}_i$ the position of atom $i$, $F_i$ the magnetic scattering tensor of atom $i$ and the sum is over all atoms. Using $\pi$ or $\sigma$ polarized incoming photons and no polarization analyzer, the two accessible channels are $\boldsymbol{\pi} \equiv I_{\pi\sigma} + I_{\pi\pi}$ and $\boldsymbol{\sigma} \equiv I_{\sigma\pi} + I_{\sigma\sigma}$, with $I_{\sigma\sigma} = 0$ for resonant magnetic scattering. The entries of the scattering tensor $F_i$ contain the orientation of the magnetic moments $\mathbf{m}_i$ and the magnetic form factors $f_i$. For the antiferromagnetic order considered in this work, $F_i$ can be simplified to

$$F_i = f_i(E)\cdot\begin{pmatrix} 0 & -\hat{m}_{i_z} & \hat{m}_{i_y} \\ \hat{m}_{i_z} & 0 & -\hat{m}_{i_x} \\ -\hat{m}_{i_y} & \hat{m}_{i_x} & 0 \end{pmatrix}.$$

Here, $f_i$ is the complex valued, energy dependent, magnetic circular dichroic form factor which we have calculated from the double cluster model described in ref. 23. For the period-four antiferromagnetic order the phases from the scattering at four subsequent sites add up to $F = F_1 + iF_2 - F_3 - iF_4$, which corresponds to $F = 2(F_{\mathrm{SB}} + iF_{\mathrm{LB}})$ in our case with $\mathbf{m}_1 = -\mathbf{m}_3$ and $\mathbf{m}_2 = -\mathbf{m}_4$ and scattering at the SB and LB sites. In principle an additional domain with exchanged SB and LB indices (i.e. exchanged small and large moment magnitudes) has to be considered, however we find experimentally that the smaller magnetic moment of the non-collinear spiral is always aligned along the [111] axis whereas the larger moment lies within the (111) plane. Our modeling includes typical 180° antiferromagnetic domains within a (111) plane, i.e. the resulting scattered intensity is averaged for example over the configurations (↑→↓←), (↑←↓→), (↓→↑←), and (↓←↑→). We then fitted the measured integrated rocking curve intensity of the (1/4, 1/4, 1/4) antiferromagnetic Bragg peak at each azimuthal angle $\psi$, simultaneously for the $\boldsymbol{\pi}$ and $\boldsymbol{\sigma}$ channels. Rotation around $\psi$ was performed with an accuracy higher than 5° in experiment. The parameters of the fit were the two angular coordinates $\Theta, \Phi$ giving the orientation of the magnetic moments and the magnitude $\delta d$ of the breathing distortion. Finally, the absolute magnetic moments sizes were obtained from the conversion of the $\delta d$ parameter by the double cluster model.[23]

**X-ray reflectivity characterization:** On- and off-resonant soft x-ray reflectivity was used to determine the layer thicknesses and interface roughnesses of $NdNiO_3$ and $NdGaO_3$ with the software ReMagX (ref. 35). The off-resonant parts of the atomic scattering factors of Nd, Ga, Ni, and O were taken from the Chantler tables.[36] The resonant parts were extracted from the corresponding x-ray absorption spectra (XAS) recorded as described in ref. 37. As an example the real and imaginary parts $f_1$ and $f_2$ of the Ni and Nd scattering factors of the $N = 55$ film are shown in the inset of Fig. 7(a) in the Supplementary Information (SI), with the real part $f_1$ obtained by Kramers-Kronig transformation. The reflectivity curves taken at various energies and $T = 300$ K were simultaneously fitted [SI-Fig. 7(a)-(d)]. The resulting set of parameters for the layer thickness $t$ and interface roughness $r$ are summarized in SI-Table I.

**Quantifying the magnetically active Ni planes:** The $NdNiO_3$ magnetic Bragg peak is observed in the resonant soft x-ray reflectivity of [111] oriented samples[27] since the $\mathbf{q}_0$=(1/4, 1/4, 1/4) propagation vector direction lies along the specular direction [SI-Fig. 7(e)-(l)]. We have used the software tool QUAD (ref. 38) to analyze the magnetic contribution to the reflectivity signal in the $2\theta$ range between 60° and 165°. Accordingly, the number $N$ of magnetically active Ni planes stacked along the [111] direction is determined (SI-Table I). The obtained numbers are consistent with the structural thickness from the independent ReMagX fits at $T$ = 300 K [SI-Fig. 7(a)-(d)], where the magnetic peak is absent. The solid lines in SI-Fig. 7(e)-(l) are the superposition of the magnetic QUAD fit and the structural ReMagX fit. As the number $N$ agrees with the structural fit and the STEM analysis, we exclude effects such as magnetic dead layers at interfaces and in particular for the thinnest sample it is deduced that effectively 8 Ni layers are active, corresponding to two full periodicities of the period-four antiferromagnetic order.

**Separation of structural and magnetic contributions:** In the $\theta$-$2\theta$ reflectivity scans in SI-Fig. 7(e)-(l) structural and magnetic contributions overlap. In order to correctly determine the variation of the magnetic scattering intensity as a function of the azimuthal angle $\psi$ [SI-Fig. 8(a),(b) and Fig. 3 main text], it is essential to avoid structural contributions to the detected intensity. This is achieved by using a narrower detector slit than for the reflectivity data shown in SI-Fig. 7, decreasing the overall intensity but increasing the experimental resolution. The finer angular acceptance of the detector then allows the magnetic peak to be distinguished from reflectivity (specular reflection) as seen in SI-Fig. 8(c). Furthermore, the intensity of specular reflection strongly decreases for high $2\theta$ values and almost drops to zero at $2\theta \sim 115°$ where the intensity of the magnetic peak is strongest. As a consequence, the structural contribution is negligible in the integrated intensities of the rocking curves at $\mathbf{q}_0 = (1/4, 1/4, 1/4)$ we analyzed in Fig. 3 of the main text. The fact that the two components are separated in the first place can be a consequence of the miscut angle of the $NdGaO_3$ substrate (here between 0.1° and 0.5°). More precisely, the substrate surface and the $NdNiO_3$ crystallographic planes (and therewith the magnetic structure) are tilted with respect to each other. Hence the specular reflection and the magnetic (1/4, 1/4, 1/4) rod do not fully coincide and the two peaks can be observed separately in a rocking curve. Finally, it is pointed out that also there are no (or very weak) interference effects between the specular reflection and the magnetic peak, as evident from SI-Fig. 8(d) and moreover the Brewster angle condition at $2\theta \sim 90°$ has no (or very small) influence on the magnetic signal [SI-Fig. 8(d)].

**Scanning transmission electron microscopy:** Cross-section specimens for scanning transmission electron microscopy (STEM) investigation were thinned to electron transparency by mechanical grinding, tripod polishing, and argon ion beam milling. Before the experiment, samples were treated in a Fischione plasma cleaner in a 75% argon - 25% oxygen mixture. The electron microscopy and spectroscopy measurements were performed on a JEOL ARM 200CF microscope equipped with a cold field-emission electron source, a probe Cs corrector, a large solid-angle SDD-type JEOL Centurio EDX detector, and a Gatan GIF Quantum ERS spectrometer. The microscope was operated at 200 kV with a semiconvergence angle of 20.4 mrad providing a probe size of less than 1 Å for the analytical analysis. Collection angles of 11-23 and 75-309 mrad were used for acquiring annular bright field (ABF) and high angle annular dark-field (HAADF) images, respectively. A collection semi-angle of 68.5 mrad was used for electron energy loss spectroscopy (EELS) measurements. For the ABF and HAADF images, frame series with short dwell times ($2\mu$s/pixel) were used and added after cross-correlation alignment to improve the signal to noise ratio. In addition, STEM images and EELS data were processed with a multivariate weighted principal component analysis (PCA) routine (MSA Plugin in Digital Micrograph) to decrease the noise level (ref. 39).

HAADF images of the $N = 8$ sample shown in panel SI-Fig. 6(a) are representative for the high structural quality of the samples with almost no grains and/or extended defects. The higher magnification image shown in the right panel of SI-Fig. 6(a) demonstrates the epitaxial layer-by-layer growth with coherent interfaces where the $NdNiO_3$ slab is discernable through the $Z$-contrast between the $NdGaO_3$ substrate and $NdGaO_3$ capping layer.

Atomic column positions and Ni-O-Ni and Ga-O-Ga bond angles were determined from the ABF images [SI-Fig. 6(b) and (c)] with the Digital Micrograph software tool described in ref. 40. The results of the Ni-O-Ni bond angle analysis of the $N = 55$ and 8 sample are shown in SI-Fig. 6.

**Density functional theory calculations:** Density functional theory (DFT) calculations were performed with the VASP (Vienna ab initio simulation package) code[41] using the generalized gradient approximation GGA-PBE functional[42] for electronic exchange and correlation on cubic (corresponding to space group $Pm\bar{3}m$), orthorhombic (corresponding to $Pbnm$), and monoclinic (corresponding to $P2_1/n$) symmetry. To capture the experimentally observed period-four magnetic order with ferromagnetic Ni planes stacked along the cubic [111] direction for bulk and different slab geometries, we defined larger, charge neutral supercells with formula units $Nd_{24}Ga_{24-N}Ni_NO_{72}$ for $N \in \{4, 8, 12, 16, 24\}$ which are commensurate with the magnetic order (SI-Fig. 9(a)). For the calculations of cubic bulk, we have first minimized the total energy by performing a full relaxation of the systems including optimization of cell volume and all individual ionic positions, conserving the symmetry group of the initial unit cell. In slab superstructures, the overall symmetry is reduced compared to the bulk due to the Ni-to-Ga ion exchange in the confining layers. The full relaxation of the slab structures included ionic positions in both sublayers, $NdNiO_3$ and $NdGaO_3$.

We employed various values of Hubbard $U$ in the GGA+$U$ calculations and subsequently compared total energies for the symmetry broken phase to determine the magnetic ground state [SI-Fig. 9(b)]. For $U$ values around 2 eV in the cubic calculation, a period-four order can be stabilized for bulk and truncated slabs as discussed in detail in the main text. For $U < 0.5$ eV and $U$ above a thickness dependent upper limit, the ferromagnetic solution is the ground state for cubic symmetry. In the bulk calculation the period-four order is stable only in a comparatively narrow $U$ range around 2 eV, but we note that a similarly small value of $U \sim 2$ eV has recently also been reported from first-principles constrained random-phase approximation studies of the rare-earth nickelates.[43]

For the monoclinic structure the DFT+$U$ calculations result in a gapped, insulating ground state with a small, but robust charge disproportionation evident from a $\lesssim \pm1\%$ variation in the $d$-projected charge around the two different Ni sites in bulk as well as truncated slabs with $N = 4, 8$. The $d$-projected charge at each site has been calculated using the VASP standard sphere and in SI-Fig. 10(a)-(c) upwards (downwards) pointing orange bars indicate the deviation in charge from the mean value. In a similar $U$ range as for the cubic case a period-four order is stable. However, depending on crystal symmetry, the DFT+$U$ results reflect two different solutions for the spin system to accommodate the nearest-neighbor spin frustration: (i) in cubic and orthorhombic ($Pbnm$; not shown) symmetry (no bond disproportionation) the frustration results in a canting of moments in the two sublattices and a stabilization of a 90° spiral in the bulk limit as discussed in the main text. (ii) For monoclinic symmetry (with bond disproportionation), the energy gain for spin polarization on the SB octahedra subsystem (Ni1 and Ni3) is not sufficient to yield an ordered moment in the bulk limit, even though the LB octahedra subsystem (Ni2 and Ni4) orders such that the ordering vector $q = (1/4, 1/4, 1/4)$ is preserved [SI-Fig 10(a)] (see also ref. 21). In truncated slabs both effects are softened at the interfaces: in the cubic case the canting disappears and in the monoclinic case the moment on the SB sublattice orders such that in both cases a collinear order is realized for $N \leq 8$ slabs [SI-Fig 10(b),(c)], however with the difference that the averaged sublattice moments are equal in the cubic case but clearly distinct in the monoclinic calculation.

It should be noted that in the calculations with lower, monoclinic symmetry atomic positions were fixed to the experimental values of $NdNiO_3$ $P2_1/n$,[25] because we found that depending on whether or not a spin polarization of the initial densities is chosen as starting condition in the monoclinic DFT+$U$ calculation, the structure relaxes either to $Pbnm$ symmetry with no bond disproportionation or retains the monoclinic symmetry and shows bond disproportionation. Whenever the monoclinic symmetry is preserved, by constraint or due to structural relaxation with initial spin polarization, a magnetic order with vanishing moment in the SB sublattice is found. In case of orthorhombic symmetry, by constraint or due to structural relaxation without initial spin polarization, the result is the experimentally observed, non-collinear state also found for cubic symmetry. In constrained monoclinic slabs, top and bottom interface layers are different, since they are either formed by SB or LB sites. This asymmetry is reflected in the site-projected $d$ charges, which are different at those two interfaces (orange bars in SI-Fig. 10). We emphasize that the ordered moment amplitudes scale independently from the charge disproportionation. Altogether we conclude that the experimentally observed coexistence of non-collinear magnetic order and static bond disproportionation in bulk $NdNiO_3$,[4,18] cannot be reproduced in the same DFT+$U$ calculation, in agreement with ref. 21. This issue has to be addressed in future work. DFT calculations with higher $k$-resolution are required to definitively resolve the relationship between the site-dependent spin polarization and the energy gap at the Fermi level.

Finally, we assessed the influence of magneto-crystalline anisotropy. To this end we included spin-orbit coupling in our supercell calculations and compared the total energies of states with easy magnetization axis along $x, y, z$ respectively, and obtained values on a sub-meV scale. This hierarchy of energy scales is generally expected for 3$d$ metals, and especially for the nickelates where the magnetic moments arise from the partially filled Ni $e_g$ orbitals with quenched orbital moment, whereas $t_{2g}$ states with nonzero $g$ factor are fully occupied. Our DFT results show that $t_{2g}$ and $e_g$ states are well separated by about 1-2 eV and bandwidths are also on the eV scale, while the spin orbit coupling of the $t_{2g}$ states of the order of 20 meV. Another possibility is magnetic anisotropy originating from non-local exchange with local moments on the rare-earth sites. Having considered the geometry of such couplings, we concluded that the projected rare-earth moments on each Ni site would lead to identical anisotropy in the global coordinate system, even in the monoclinic phase. Altogether, we can rule out any significant influence of the spin-orbit coupling and the magneto-crystalline anisotropy on the observed collinear-to-noncollinear transition.

**Data availability Statement**

The data that support the plots within this paper and other findings of this study are available from the corresponding author upon reasonable request.

# Supplementary Information for
# "Complex magnetic order in nickelate slabs"

M. Hepting, R. J. Green, Z. Zhong, M. Bluschke, Y. E. Suyolcu, S. Macke, A. Frano, S. Catalano, M. Gibert, R. Sutarto, F. He, G. Cristiani, G. Logvenov, Y. Wang, P. A. van Aken, P. Hansmann, M. Le Tacon, J.-M. Triscone, G. A. Sawatzky, B. Keimer, and E. Benckiser*

**Contents**

*E.Benckiser@fkf.mpg.de

## A. Figures and Tables referenced in the Methods section

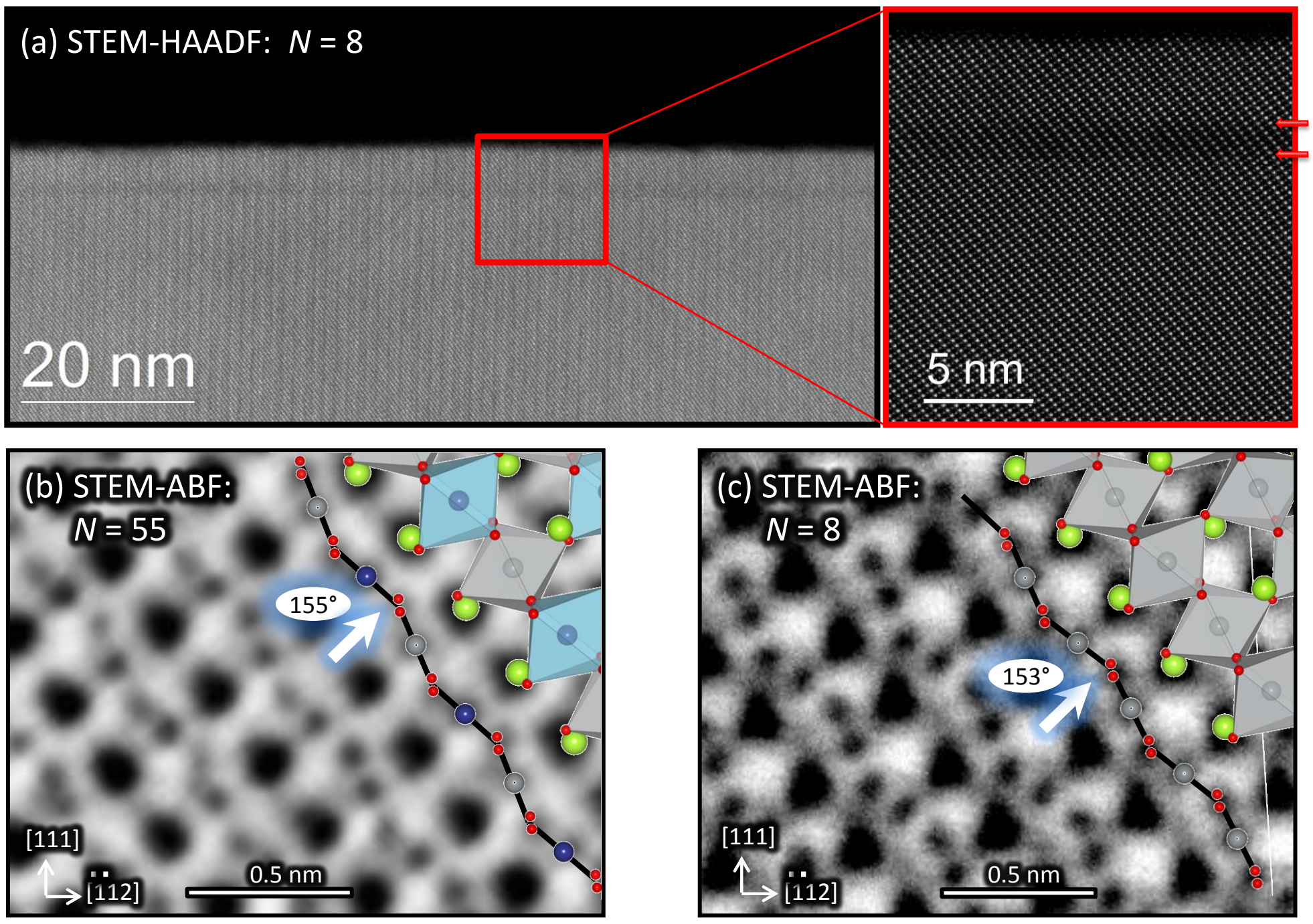

Fig. 6: **STEM analysis of the $N = 55$ and 8 slab samples.** (a) Low magnification (left) and high magnification (right) STEM-HAADF images of the $N = 8$ sample. The images show coherent interfaces where the $NdNiO_3$ slab is discernable through $Z$-contrast between the $NdGaO_3$ substrate and $NdGaO_3$ capping layer. The red arrows mark the nominal interfaces in the high magnification image. (b),(c) ABF images mapping the oxygen positions in $N = 55$ and $N = 8$ slab, respectively. A quantitative analysis of the Ni-O-Ni bond angles reveals a subtle relaxation in the $N = 55$ case ($\sim 155°$), whereas the oxygen positions in the $N = 8$ case ($\sim 153°$) are pinned to the confining $NdGaO_3$ oxygen octahedron network. The error bars of the bond angle determination (95% confidence interval, corresponding to 2 times the standard error) are less than 1° (ref. 1).

Table I: **Structural and magnetic parameters extracted from x-ray reflectivity.** Layer thickness $t$ and interface roughnesses $r_1$ ($NdGaO_3$ substrate to $NdNiO_3$) and $r_2$ ($NdNiO_3$ to $NdGaO_3$ capping) are obtained from fits to on -and off resonant scans at $T = 300$ K. The error of $t$ is 1-2 Å and less than 0.5 Å for $r$. The number $N$ of magnetically active Ni planes in the [111] direction is deduced from the modeling of the magnetic contribution in the resonant scans at $E = 853$ eV and $T = 20$ K and the errors of $N$ are extracted from the analysis shown Fig. 7.

| sample | $t$ $NdNiO_3$ (Å) | $t$ $NdGaO_3$ (Å) | $r_1$ (Å) | $r_2$ (Å) | $N$ magnetic Ni planes |
|---|---|---|---|---|---|
| 55 Ni | 113.9 | – | 3.6 | – | 55 ± 3 |
| 45 Ni | 97.6 | – | 3.2 | – | 45 ± 3 |
| 14 Ni | 26.1 | 50.4 | 3.3 | 4.1 | 14 ± 1 |
| 8 Ni | 14.4 | 48.9 | 3.2 | 5.4 | 8 ± 0 |

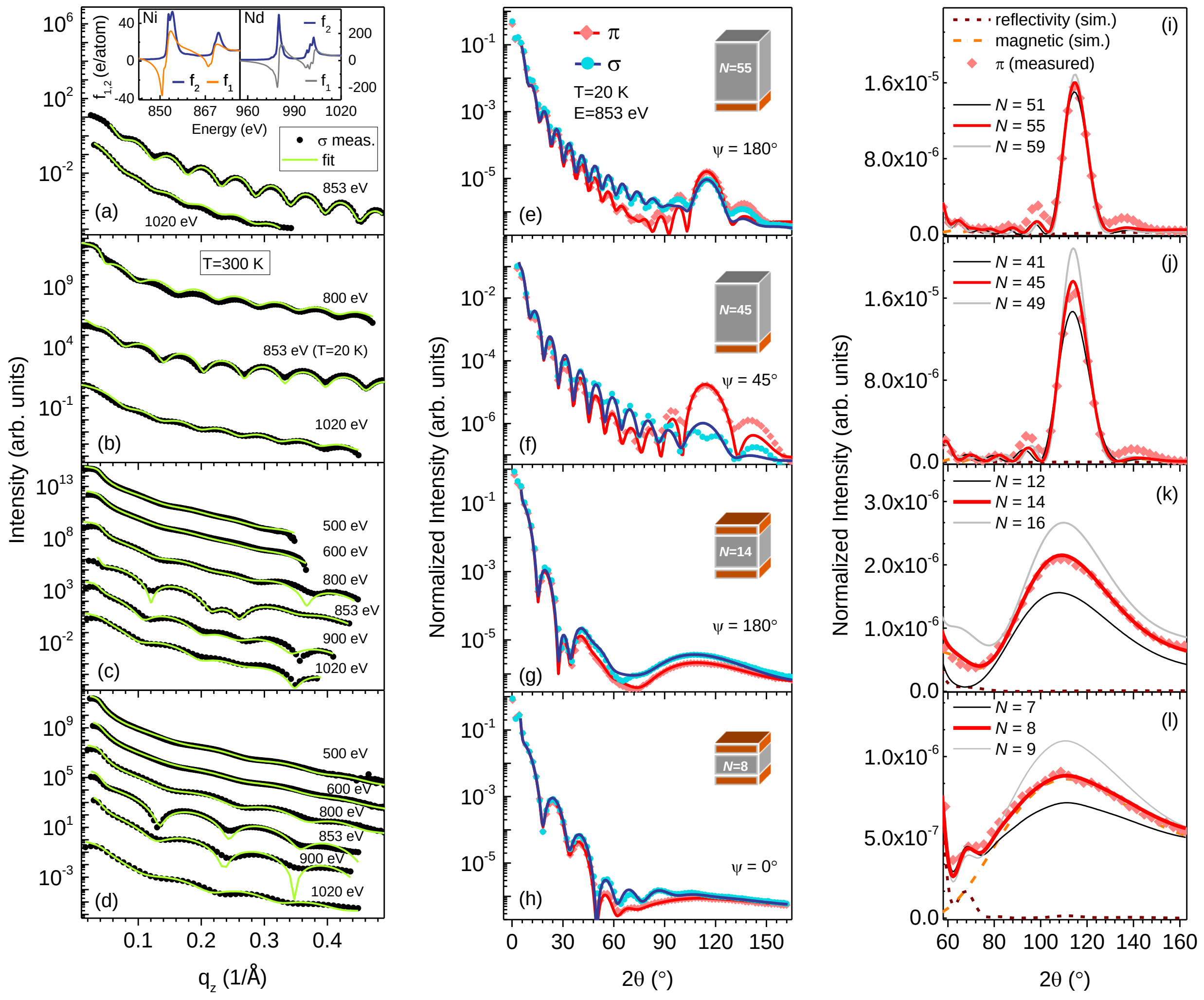

Fig. 7: **Characterization of structural and magnetic properties.** (a)-(d) Reflectivity of the $N = 55$, 45, 14, and 8 Ni layers at various energies at $T = 300$ K (black symbols). The $NdNiO_3$ film thickness and the $NdGaO_3$ capping layer thickness are extracted from the fits (solid green lines). For clarity the reflectivity curves at different energies are offset. The inset in panel (a) shows the imaginary and real part ($f_2$ and $f_1$) of the Ni and Nd scattering factors, exemplary for $N = 55$ sample. (e)-(h) Resonant reflectivity at $E = 853$ eV (Ni $L_3$) and $T = 20$ K for $\pi$ and $\sigma$ polarized photons (red and blue symbols). The solid lines are the superposition of the structural fit at $E = 853$ eV from panel (a)-(d) and separate simulations to the antiferromagnetic (1/4, 1/4, 1/4) peak appearing in the reflectivity at $2\theta \sim 115°$. (i)-(l) Enlarged view of the antiferromagnetic peak in (e)-(h) on a linear intensity scale (red symbols) together with simulated curves benchmarking the error bars for the number of magnetically active Ni planes (black and gray lines). The red solid line in panels (i)-(l) is the simulation which agrees best with the measurement data. Accordingly, the results for the number of magnetically active Ni layers are $N = 55 \pm 3$, $45 \pm 3$, $14 \pm 1$, and $8 \pm 0$.

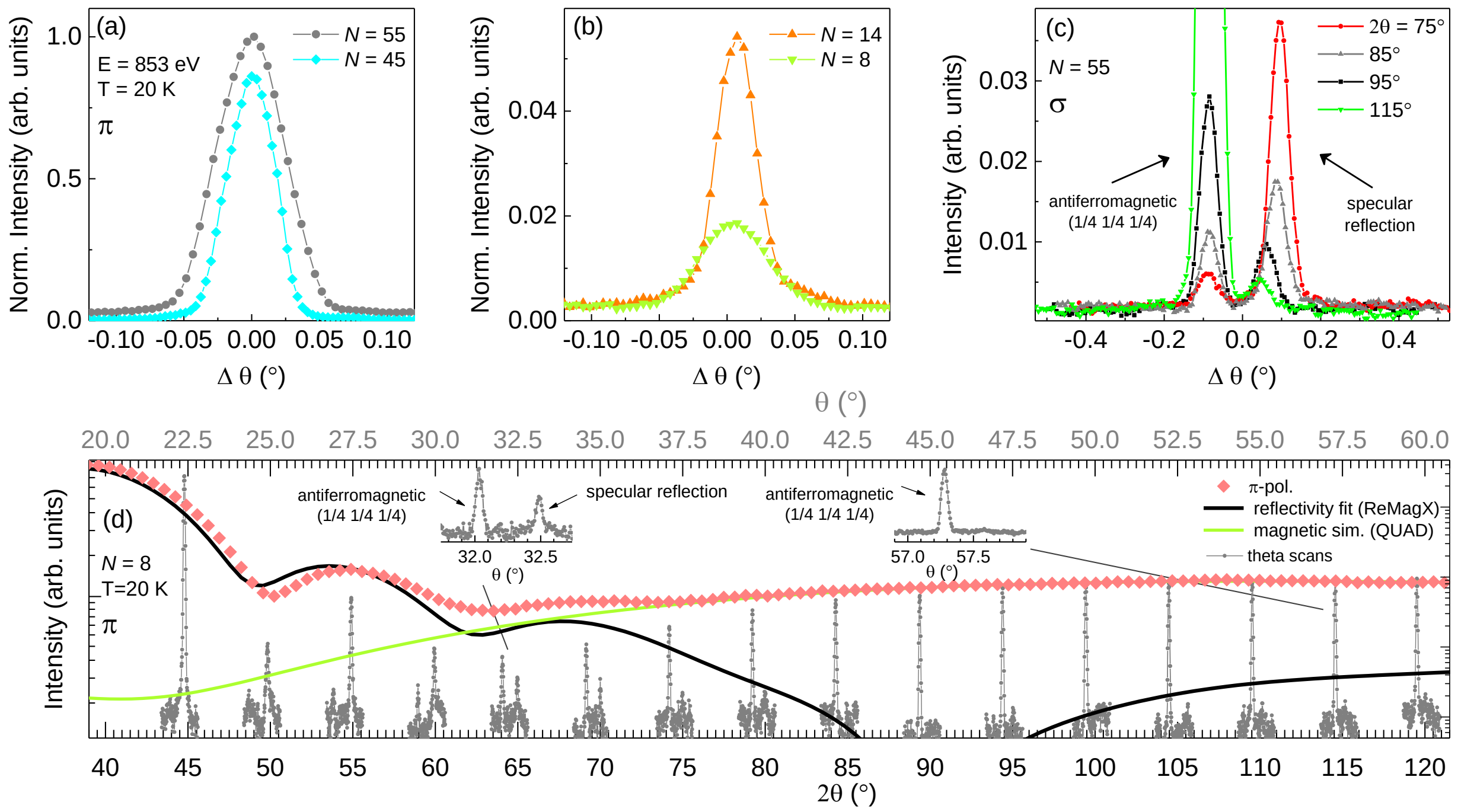

Fig. 8: **Antiferromagnetic Bragg peak and separation of magnetic and structural contributions.** (a),(b) Comparison of the antiferromagnetic Bragg reflection at $\mathbf{q}_0 = (1/4,\ 1/4,\ 1/4)$ of the $N = 55$, 45, 14, and 8 Ni layers in a rocking curve recorded with a narrow detector slit. The $N = 55$ peak intensity is normalized to unity and the other curves are scaled accordingly. (c) Separation of peaks with magnetic and structural origin at different $2\theta$ angles. The magnetic peak reaches its maximum intensity at $2\theta \sim 115°$ whereas the intensity coming from the specular crystal truncation rod drops to almost zero for high $2\theta$ angles. In consequence, our analysis of rocking curves of the magnetic peak is free from structural contributions. (d) Detailed evolution of the structural and magnetic components. Red symbols are the reflectivity recorded with a wide slit. The solid black line is the ReMagX[2] fit to the structural part of the reflectivity, the solid green line is the QUAD[3] simulation of the magnetic peak contribution. Additionally, rocking curves (gray) were taken with a narrow slit at various $2\theta$ angles. The intensity of the peak at smaller $\theta$ follows the magnetic simulation (green line) whereas the intensity of the peak at higher $\theta$ follows the fit of the reflectivity (black line). The observation that the green and the black curve, obtained from two independent fits, serve as envelopes for the two peaks in the rocking scans indicates that there is no (or very small) interference between the magnetic and the structural components in our data.

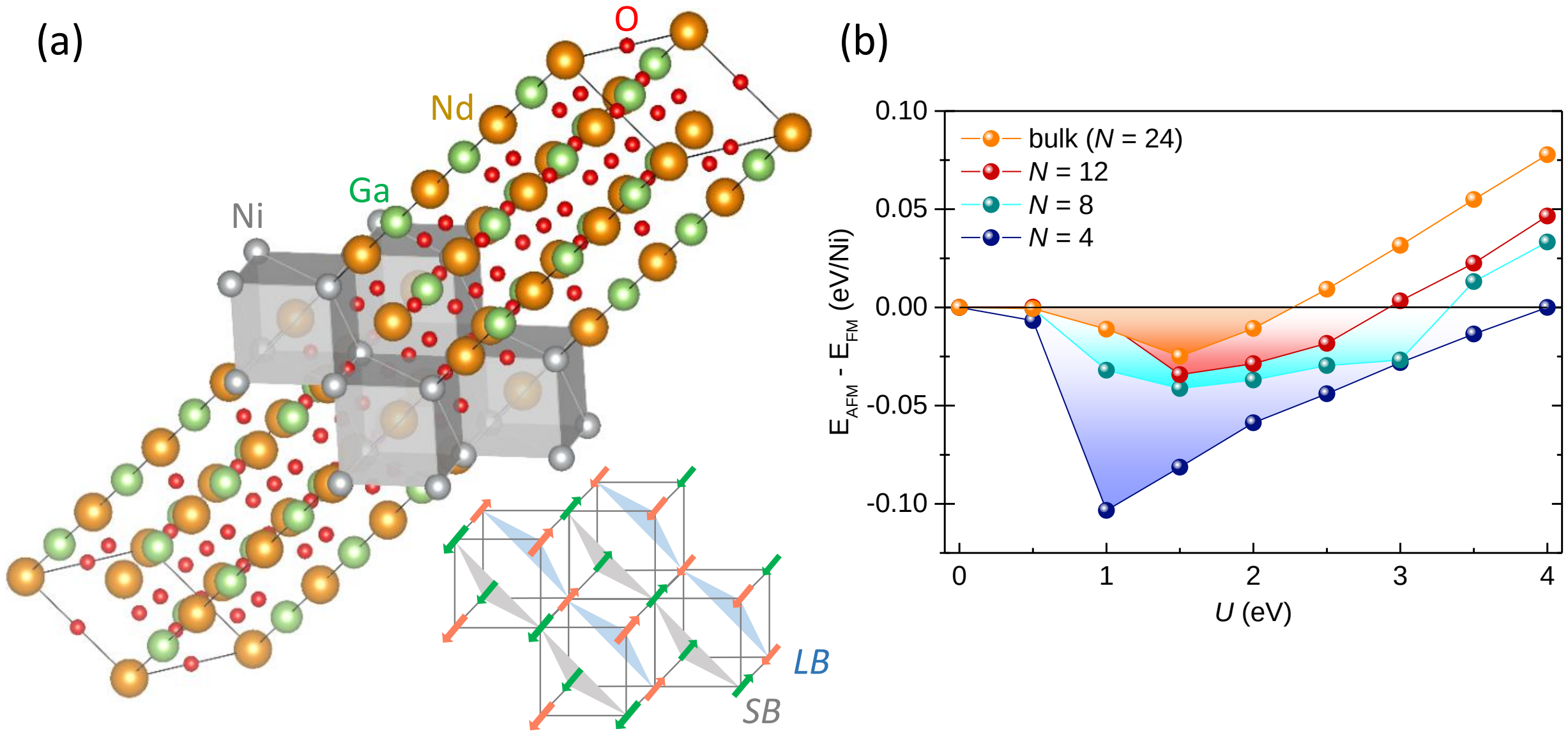

Fig. 9: **DFT magnetic ground state for cubic structures.** (a) Representation of the $Nd_{24}Ga_{24-N}Ni_NO_{72}$ supercell,[4] here for $N = 4$, considered in the DFT+$U$ calculations together with a corresponding sketch of alternating, ferromagnetic SB and LB planes along the pseudo-cubic [111] direction. (b) Comparison of total energies of antiferromagnetic configurations with the corresponding ferromagnetic state as a function of the effective $U$ parameter for calculations performed in supercells as sketched in (a) with relaxed atomic positions. Around $U \sim 2$ eV we find a minimum for the antiferromagnetic period-four magnetic order, where a non-collinear state is realized for the bulk and a collinear state is found to be the ground state for $N = 4$ and 8 Ni planes.

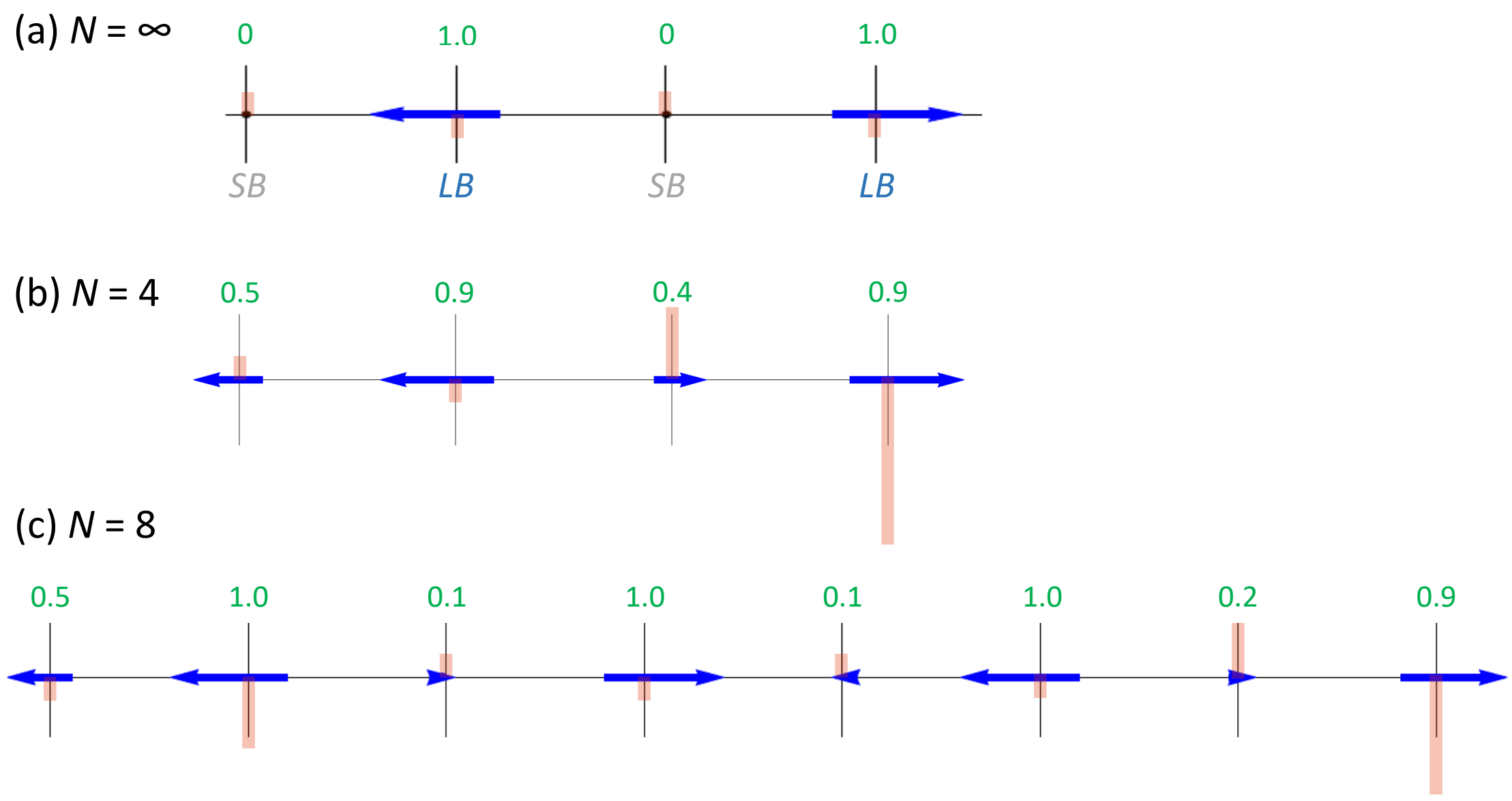

Fig. 10: **DFT+$U$ magnetic ground states for monoclinic symmetry** (a)-(c) Representation of the relative orientation and amplitude of ordered magnetic moments obtained from integration of the site- and spin-resolved density of states for $U = 2$ eV and monoclinic crystal symmetry. (a) The unrestricted $N = 24$ supercell corresponding to bulk $NdNiO_3$ converges to a period-four magnetic ground state where the energy gain for spin polarization on the SB octahedra subsystem (Ni1 and Ni3) is not sufficient to yield an ordered moment. (b), (c) In truncated slabs with $N \leq 8$ Ni layers a nonzero ordered moment at the Ni1 and Ni3 sites appears such that a collinear ↑↑↓↓-type of order is realized. For bulk and finite slab structures the DFT+$U$ results show a small, but stable charge disproportionation which is reflected in a site-dependent variation of $d$-projected charge illustrated by the orange lines in each panel (see text).

### B. Orbital polarization

The x-ray absorption spectra (XAS) in Fig. 2(a) of the main text are measured on a wedge with an inclination angle of 54.7° together at an azimuthal angle of 45°. In this scattering geometry the two different orientations of the electric field vectors of the incoming x-rays correspond to the spatial extension of the lobes of the Ni $e_g$ orbitals with $x^2$-$y^2$ and $3z^2$-$r^2$ character, respectively. The vanishing linear dichroism $\sigma$-$\pi$ and the hole ratio of $X \sim 1$ indicates the degeneracy of the $d_{3z^2-r^2}$ and $d_{x^2-y^2}$ orbitals and the absence of orbital polarization influencing the spin order.[5] This is in accordance with the expectation that a trigonal lattice distortion (here due to tensile strain in the (111) plane) does not lift the degeneracy of $e_g$ orbitals.[6] We note that ref. 7 reported a non-zero dichroic XAS signal in a comparable scattering geometry for much thinner $N = 2$ samples of $NdNiO_3$-$LaAlO_3$ which was interpreted as signature of antiferro-orbital order. The XAS shown in Fig. 11(a)-(c) are measured in a flat scattering geometry without wedge, i.e. a quantitative extraction of a possible imbalance in the $d_{3z^2-r^2}$ and $d_{x^2-y^2}$ occupation is more complex than the formalism given in ref. 8, since the polarization vector directions of the incoming photons and the spatial elongations of the orbitals do not directly coincide. However, the difference signals $\boldsymbol{\sigma} - \boldsymbol{\pi}$ of the $N = 55$, 14 and 8 Ni planes samples is zero within the noise limit [Fig. 11(a)-(c)], thus we exclude the emergence of an orbital polarization through truncation along the [111] direction for all our samples.

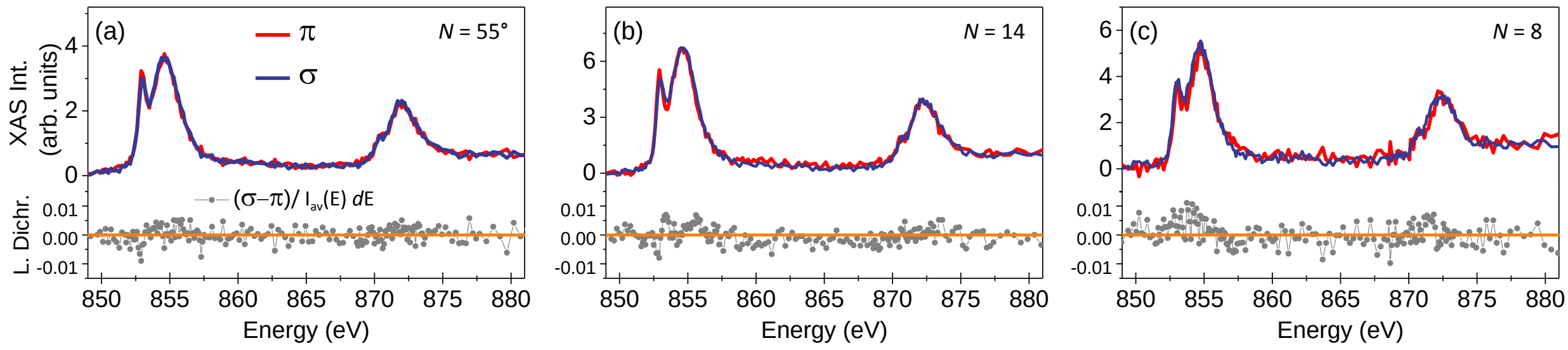

Fig. 11: **Absence of orbital polarization and its influence on the spin order.** (a)-(c) XAS of the $N = 55$, 14, and 8 Ni layers sample at $T = 20$ K at the Ni $L_{3,2}$ edge for $\pi$ and $\sigma$ polarized incoming photons, measured in horizontal scattering geometry in fluorescence yield mode. The difference spectrum $\boldsymbol{\sigma} - \boldsymbol{\pi}$ normalized by the averaged energy integral from 850 eV to 880 eV is zero within the noise for all samples, i.e. there is no significant change from $N = 55$ to $N = 8$ indicating the emergence of an orbital polarization.

### C. Alternative magnetic model fits

Assuming two spin sublattices with moments $\mathbf{m}_1$ and $\mathbf{m}_2$, a certain projection of the moments not only follows from one unique spin configuration but also after specific reorientations of the spins combined with a change in the imbalance between the $\mathbf{m}_1$ and $\mathbf{m}_2$ magnitudes.[5,9] To account for this we have defined the following strategy for the fitting procedure: first, we use the parameters $\Theta, \Phi$, and $\delta d$ of the orthogonal (↑→↓←) bulk model[9,10] (Table II) as starting values and vary the parameters until convergence with the experimental azimuthal dependence is reached. The results of 'free fit 1' are summarized in Table II and illustrated in Fig. 3(h)-(k) of the main text. Second, we use a large set of random starting values and let the fit converge. Most sets of starting parameters yield the same result as 'free fit 1' Table II. For the $N = 55$ and 14 data there is basically one alternative solution ('free fit 2') with a similarly small least $\chi^2$ (Table III). Note that the 45 Ni sample behaves qualitatively similar to the 55 Ni layers, hence it will not be discussed further. For the $N = 8$ dataset all fits converge to 'free fit 1', regardless of the starting parameters. Third, we fix $|\mathbf{m}_1| = |\mathbf{m}_2|$, which corresponds to previous works considering only equivalent magnetic Ni sites[5,9,10] or the presence of two equally populated domains with $|\mathbf{m}_1| > |\mathbf{m}_2|$ and $|\mathbf{m}_1| < |\mathbf{m}_2|$, respectively. In this type of modeling the form factor correction is obsolete. The obtained fit parameters are summarized in Table IV. Fourth, we use the same modeling tool as the authors Frano *et al.* in ref. 5 (no form factor correction) but allow a linear scaling factor between the scattering from two inequivalent Ni sites. Results are summarized as 'free fit' 1 and 2 in Table IV.

As the final step of our strategy we evaluate the four approaches against each other. Starting with 'free fit' 1 and 2 from Table IV we notice the close similarity of the angles $\Theta$ and $\Phi$ as compared to 'free fit' 1 and 2 from the form factor corrected fits in Tables II and III, illustrating the robustness of the solutions concerning the angular orientation of the spins. The $|\mathbf{m}_2|/|\mathbf{m}_1|$ ratio of 4.12 and 1.95, however, appears to be largely overestimated, considering that the highly monoclinic $HoNiO_3$ exhibits a $|\mathbf{m}_2|/|\mathbf{m}_1|$ ratio of only 2.33 (ref. 11). The $|\mathbf{m}_1| = |\mathbf{m}_2|$ solution of Table IV is also excluded as the energy dependence of the $\pi/\sigma$ ratio in Fig. 5 of the main text indicates the presence of two significantly different Ni scattering sites. Remaining are 'free fit' 1 and 2 in the form factor corrected model (Table II and Table III) with similar $\chi^2$ values. However, regarding the spin angles and the $\delta$d of bulk-like $NdNiO_3$ (0.037 Å), the solution with the parameters of Table II are matching closer and moreover provide an almost gradual evolution from (↑→↓←) to (↑↑↓↓) with increasing [111] truncation. The slightly higher value of $\delta$d = 0.041 Å in Table II is in agreement with (ref. 13) which locates [111] oriented $NdNiO_3$ films at the position of $SmNiO_3$ in the bulk nickelate phase diagram but does not exceed $\delta$d = 0.048 Å of the highly distorted $HoNiO_3$.

Table II: **Fit parameters in the form factor corrected model.** Summary of parameters obtained from the form factor corrected model fits and illustrated in Fig. 3 (g)-(k) of the main text. $\Theta$ is the angle between the spin vector and the [111] axis, $\Phi$ the angle to the [11$\bar{2}$] axis in a right-handed system. Indices 1, 2 indicate the first and second spin of the period-4 order, the remaining two spins are the corresponding negative vectors. For comparison the parameters of bulk $NdNiO_3$ (refs. 9,12) are listed. The errors of $\Theta_{1,2}, \Phi_{1,2}$, and $\delta d$ are interdependent and estimated to lie between 3° and 5° for $\Theta$ and $\Phi$, and $\sim 0.002$ Å for $\delta d$, as evaluated from similar least $\chi^2$ values of the fits.

| sample | $\Theta_1$ (°) | $\Phi_1$ (°) | $\Theta_2$ (°) | $\Phi_2$ (°) | $\delta d$ (Å) | $\chi^2$ |
|---|---|---|---|---|---|---|
| 55 Ni | 5 | 176 | 80 | 39 | 0.041 | 1.19E-2 |
| 45 Ni | 6 | 175 | 76 | 78 | 0.042 | 2.14E-2 |
| 14 Ni | 9 | 179 | 73 | 87 | 0.019 | 6.81E-4 |
| 8 Ni | 8 | 177 | 5 | 178 | 0.009 | 8.69E-5 |
| bulk $NdNiO_3$ | 0 | 180 | 90 | 0 | 0.037 | – |

Table III: **Alternative fit parameters in the form factor corrected model.** Starting from a random set of parameters $\Theta, \Phi$ and $\delta d$, the fit either converges to "free fit 1" (Table II) or "free fit 2". For the sample having 8 Ni layers no alternative spin orientation is listed since all fits converged to the collinear solution shown in Table II (within the error bars).

| free fit 2 | $\Theta_1$ (°) | $\Phi_1$ (°) | $\Theta_2$ (°) | $\Phi_2$ (°) | $\delta d$ (Å) | $\chi^2$ |
|---|---|---|---|---|---|---|
| 55 Ni | 17 | 53 | 80 | 39 | 0.049 | 1.19E-2 |
| 14 Ni | 41 | -113 | 113 | -94 | 0.025 | 8.89E-4 |

None of our fits converged to a spin configuration with one magnetic sublattice carrying zero moment, corresponding to the (↑ 0 ↓ 0) model, which is widely used as starting point for theory calculations and has been discussed as one possible solution of the neutron powder diffraction refinements (see references in the main text). We always find two distinct magnetic sublattices, with both moments $\mathbf{m}_1$, $\mathbf{m}_2$ being non-zero. This is particularly evident in the $N = 14$ case of Fig. 3(d) of the main text where an asymmetry in the modulation amplitude with respect to $\psi$ smaller and larger than 180° is present. Such an asymmetry can only be generated from two unequal but non-zero sublattices. However, we note that in a fully collinear case, our data analysis is not sensitive to the magnitude of the second spin lattice. The fit of the $N = 8$ data describes two slightly different sublattices with an angle of $\sim 3°$ between sublattice moments, i.e. the sublattices are effectively collinear. Taking also into account the systematics in the reduction of the breathing distortion from $N = 55$ to 14 and the tendency of an equalization of moments, we conclude that for the $N = 8$ case a configuration of $(\uparrow\uparrow\downarrow\downarrow)$ instead of $(\uparrow 0 \downarrow 0)$ is realized. In addition, the $(\uparrow 0 \downarrow 0)$ configuration corresponds to the extreme limit of the bond disproportionation with 2 Ni $d^8\underline{L} \rightarrow$ Ni $(d^8\underline{L}^2)_{S=0}$ + Ni $(d^8)_{S=1}$, with $\underline{L}$ denoting a ligand hole at the oxygen and $S$ is the total spin, i.e. one set of octahedra carries the full moment whereas the other set is fully screened by the two ligand holes. To reproduce such a disproportionation an unrealistic breathing distortion of $\delta$d $\gg 0.2$ Å has to be assumed in double cluster model calculations along the lines of ref. 14. Furthermore, the STEM data (Fig. 6) indicate that the breathing distortion in the $N = 8$ slab is vanishing due to the structural pinning.

Table IV: **Alternative fit parameters without form factor correction.** Model fit without form factor correction in analogy to ref. 5. First the magnitudes of the magnetic sublattices are fixed to be identical, $|\mathbf{m}_1| = |\mathbf{m}_2|$. Then a simple scaling factor for the magnetic scattering at the two different Ni sites is allowed ("free fit 1,2").

| $\lvert\mathbf{m}_1\rvert = \lvert\mathbf{m}_2\rvert$ | $\Theta_1$ (°) | $\Phi_1$ (°) | $\Theta_2$ (°) | $\Phi_2$ (°) | $\lvert\mathbf{m}_2\rvert/\lvert\mathbf{m}_1\rvert$ | $\chi^2$ |
|---|---|---|---|---|---|---|
| 55 Ni | 113 | -138 | 92 | 37 | 1 | 1.13E-2 |
| 14 Ni | 46 | 98 | 94 | 83 | 1 | 5.23E-4 |
| free fit 1 | $\Theta_1$ (°) | $\Phi_1$ (°) | $\Theta_2$ (°) | $\Phi_2$ (°) | $\lvert\mathbf{m}_2\rvert/\lvert\mathbf{m}_1\rvert$ | $\chi^2$ |
| 55 Ni | 6 | 174 | 80 | 40 | 4.12 | 1.13E-2 |
| 14 Ni | 9 | 179 | 73 | 87 | 1.95 | 6.80E-4 |
| free fit 2 | $\Theta_1$ (°) | $\Phi_1$ (°) | $\Theta_2$ (°) | $\Phi_2$ (°) | $\lvert\mathbf{m}_2\rvert/\lvert\mathbf{m}_1\rvert$ | $\chi^2$ |
| 55 Ni | 17 | 53 | 81 | 39 | 4.42 | 1.12E-2 |
| 14 Ni | 41 | -114 | 113 | -95 | 2.01 | 7.78E-4 |

### D. Azimuthal dependence in thick $NdNiO_3$ films

To date, no large single crystals of $NdNiO_3$ are available. In order to obtain a bulk reference for the azimuthal dependence of the magnetic scattering intensity, we performed measurements on two, very thick $NdNiO_3$ films grown by magnetron sputtering, namely a 72 nm ($N \sim 350$) and a 100 nm ($N \sim 480$) thick film. In Fig. 12 the data for individual $\sigma$ and $\pi$ polarizations, panel

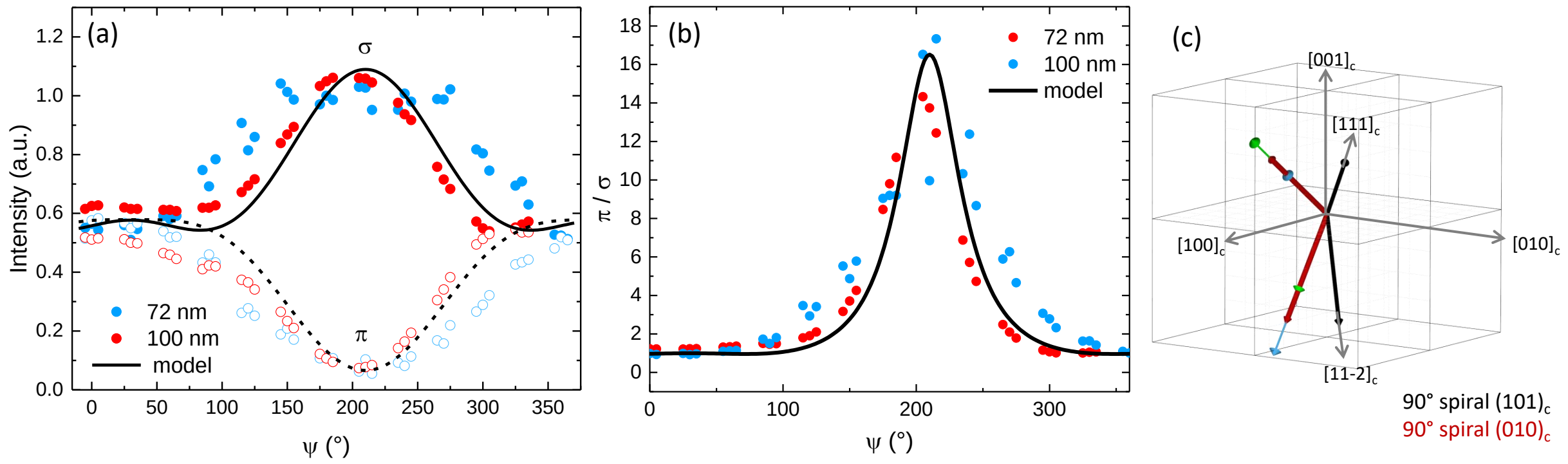

Fig. 12: **Azimuthal dependence of the magnetic scattering of thick, bulk-like $NdNiO_3$ films.** (a) Experimental azimuthal angle ($\psi$) dependence of Ni $L_3$ resonant magnetic scattering for individual $\sigma$ (solid circles) and $\pi$ (open circles) channels. (b) Corresponding $\sigma/\pi$ intensity ratios. In both panels, black lines show the best model discussed in the text. (c) Sublattice moment directions within the cubic frame corresponding to the model shown in (a,b) (red arrows: no disproportionation, green and blue arrows: bond-ordered domains with swapped LB and SB sites). For comparison the moment directions of the orthogonal non-collinear order with moments in the cubic (101) plane proposed in ref. 9 is shown (black arrows).

(a), are shown together with the corresponding $\sigma/\pi$ ratios in panel (b). The sets of data for the two films are similar, but clearly distinct from the data of the $N = 55$ Ni layer slab shown in the main text and also from a previously reported ∼50 nm thick film.[9] The most obvious difference is the $2\pi$-azimuthal periodicity of the data. The model that reproduces our data best describes a non-collinear order with moments in the $(010)_{\text{cubic}}$ plane (Fig. 12(c)).[15] When assuming a single domain the best fit is obtained with vanishing bond disproportionation, seemingly contradicting previous findings. We relate this to a complex domain structure present in very thick, relaxed films, which has not been studied in greater detail. For example, the superposition of two bond-order domains with swapped (LB, SB)-sites and nonzero bond disproportionation ($\Delta\delta = 0.037$ Å in Fig. 12(c)), both contributing equally to the magnetic scattering at the same $q_{\text{mag}}$ would yield the identical azimuthal dependence as the model shown in Fig. 12(a),(b)). Although an unambiguous determination of the moment directions and amplitudes in bulk-like thick films remains elusive, the presence of a fully *orthogonal*, non-collinear period-four order is a very robust finding. Deviations from the 90° mutual spin orientation of more than 15° resulted in clearly distinct azimuthal dependencies in all tested models.

---

15. A solution with moments in the $(100)_{cubic}$ plane is equally possible since the orientation of the samples was known only up to a 90° in-plane rotation and except from the $\phi$ shift both solutions yield the same azimuthal dependence.